\documentclass[12pt,reqno]{amsart}
\usepackage{epsfig}
\usepackage{graphicx}
\begin{document}
\baselineskip=0.8cm 
\theoremstyle{plain}
\newtheorem{thm}{Theorem}[section]
\newtheorem{lem}{Lemma}[section]
\newtheorem{prop}{Proposition}[section]
\newtheorem{coll}{Conclusion}
\theoremstyle{remark}
\newtheorem{rem}{Remark}[section]
\title{Nonsingular positon and complexiton solutions for the coupled KdV system}
\author{H. C. Hu$^{1,2}$, Bin Tong$^{1}$ and S. Y. Lou$^{1, 3}$}
\dedicatory{$^1$Department of Physics, Shanghai Jiao Tong
University, Shanghai, 200030, P. R. China \\
$^2$ College of Science, University of Shanghai for Science and Technology, Shanghai, 200093, P. R. China \\
$^3$ Department of Physics, Ningbo University, Ningbo, 315211, P.
R. China}
\begin{abstract}
Taking the coupled KdV system as a simple example, analytical and
nonsingular complexiton solutions are firstly discovered in this
letter for integrable systems. Additionally, the analytical and
nonsingular positon-negaton interaction solutions are also firstly
found for S-integrable model. The new analytical positon, negaton
and complexiton solutions of the coupled KdV system are given out
both analytically and graphically by means of the iterative
Darboux transformations.

 \vskip.2in \leftline{\bf \emph{PACS.}02.30.Ik, 02.30.Jr,
05.45.Yv.}
\end{abstract}
\maketitle

\section{Introduction}

On the exact solutions of integrable models, there is a new
classification way recently based on the property of the spectral
parameter \cite{Ma}: If the spectral parameter is positive, the
related solution of the nonlinear system is called positon which
is usually expressed by means of the triangle functions. If the
spectral parameter is negtive, the related solution of the
nonlinear system is called negaton which is usually expressed with
help of the hyperbolic functions. The term ``positon" and
``negaton" may be pursued back to Matveev et al \cite{Matveev,CUA}
though the singular solutions had been studied earlier \cite{MD}.
The so-called complexiton, which is expressed by combinations of
the triangle functions and the hyperbolic functions, is related to
the complex spectral parameters. It is meaningful that the
complexiton exists for \em real \rm integrable systems. The most
of known positon solutions are singular. For various important
integrable systems such as the KdV equation there is no
nonsingular positon found. The negaton can be both singular and
nonsingular. Especially nonsingular negatons are just the usual
solitons. The interaction solutions among positons and negatons
have also been studied recently\cite{LianLou,YChen}.
Unfortunately, except for a special C-integrable model, the STO
(Sharma-Tass-Olver) model \cite{LianLou}, in our knowledge, there
is no known nonsingular positon-negaton interaction solution.
Though the complexiton is proposed for several years, but up to
now, one has not yet found a nonsingular complexiton solution for
any (1+1)-dimensional integrable systems. In high dimensions,
(2+1)- and (3+1)-dimensions, some special types of analytic
nonsingular positon, posilton-negaton interaction solutions and
the complexiton solutions can be easily obtained because of the
existence of the arbitrary functions in their expressions of exact
solutions\cite{LouHu,LouTang}.

Then an important and interesting question is naturally arised: \\
\em Are there any nonsingular complexiton and positon-negaton
solutions for S-integrable models? \rm
  As we all know, the Korteweg-de Vries (KdV) equation is a
  typical and classical model to describe the weakly shallow long
  waves in the soliton theory. Many important properties of KdV
  equation, such as infinitely many symmetries, infinitely many conservation laws \cite{MGK},
  inverse scattering transform,
  B\"acklund transformation, Darboux transformation, etc. have been
  studied extensively for a long time and it can be viewed as a
  completely integrable Hamiltonian system \cite{Magri, Zakharov}.
Nevertheless, it is known that both the positon and complexiton
solutions of the KdV equation are singuler\cite{Matveev,Ma}.

On the other hand, coupled integrable systems, which come up in
many physical fields,  have also been studied extensively  and a
lot of interesting results are given out from both the
classification view and application field \cite{Vladimir}. The
first coupled KdV system was put forward by Hirota and Satsuma
(HS) in 1981 \cite{Hirota}. The multi-soliton solutions,
conservation laws and Darboux transformation, etc for the HS model
are studied in detail \cite{Hirota1}. Many other coupled KdV
systems such as the Ito's system \cite{Ito}, the Drinfeld and
Sokolov \cite{Sokolov} model, the Fuchssteiner equation
\cite{fush}, the Nutk-O\~guz model\cite{NO}, the Zharkov system
\cite{Zharkov} and the Foursov model \cite{Foursov} are also
constructed. Vladimir gives a classification of the integrable
coupled KdV systems by using a symmetry approach \cite{Vladimir}.

In the real application aspect of the coupled KdV systems,  a
quite general coupled KdV equation,
\begin{eqnarray}
&&u_{t}-6uu_x+u_{xxx}+\epsilon_1vv_x+\epsilon_2(uv)_x+\epsilon_3v_{xxx}=0,\nonumber\\
&&v_t-6\beta vv_x+\beta
v_{xxx}-V_0v_x+\alpha[\epsilon_1(vu)_x+\epsilon_2uu_x+\epsilon_3u_{xxx}]=0\label{ckdv1}
\end{eqnarray}
where $\epsilon_1,\ \epsilon_2,\ \epsilon_3,\ \beta$ and $V_0$ are
arbitrary constants, is derived from a two-wave modes in a shallow
stratified liquid \cite{CKDV}.

Another type of general coupled KdV system
\begin{eqnarray}
&&u_{t}+\alpha_1vu_{x}+(\alpha_2v^2+\alpha_3uv+\alpha_4u_{xx}+\alpha_5u^2)_x=0,\nonumber\\
&&v_{t}+\delta_1vu_{x}+(\delta_2u^2+\delta_3uv+\delta_4v_{xx}+\delta_5v^2)_x=0,\label{ckdv2}
\end{eqnarray}
with arbitrary constants $\alpha_i,\ \delta_i,\ i=1,\ 2,\ ...,\ 5$
is obtained from a two layer model of the atmospheric dynamical
system \cite{LTHT}.

A more general coupled KdV system
\begin{eqnarray}
\partial_\tau q_j+\partial_\zeta \sum_{k,l}\gamma_j^{kl}q_kq_l
+\partial_\zeta^3\sum_{k}\beta_j^kq_k=0,\ j=1,\ 2\label{ckdv3}
\end{eqnarray}
is derived to describe two-component Bose-Einstein condensates
\cite{BK}.

In this Letter, we study the exact solutions, especially the
positons, negatons and complexitons of a special coupled KdV
system
\begin{equation}
u_t+6vv_x-6uu_x+u_{xxx}=0,\qquad
v_t-6uv_x-6vu_x+v_{xxx}=0.\label{ckdv}
\end{equation}
It is clear that the coupled KdV system \eqref{ckdv} is a special
case for all the physical systems \eqref{ckdv1}, \eqref{ckdv2} and
\eqref{ckdv3}. It is also interesting that, the system
\eqref{ckdv} can be read off simply from the real and imaginary
parts of the single complex KdV equation
\begin{equation}\label{kdv}
U_t-6UU_x+U_{xxx}=0,
\end{equation}
where
$$U=u+I v,\ I\equiv \sqrt{-1}.$$

From the fact that \eqref{ckdv} can be derived from the complex
KdV equation, one can reasonable believe that all the integrable
properties of the real KdV equation can be preserved. Actually, we
have really proved its Painlev\'e property though we haven't write
down the proof procedure here because the method is standard and
even can be performed simply by pressing the ``enter" key with
help of some existent Maple or Mathematica programmes related to
the Painlev\'e test, say, ``Ptest" by Xu and Li \cite{XuLi}.

In the remained part of this paper, we use the Darboux
transformation which is one of the most powerful tools in
constructing exact solutions for many integrable nonlinear
equations to the coupled KdV system \eqref{ckdv} to study its
positon, negaton, complexiton solutions and the interaction
property among these types of excitations.

\section{Lax pair and Darboux transformations of the coupled KdV system \eqref{ckdv}.}

  In this section, we apply the Darboux transformation
  to the coupled KdV system \eqref{ckdv} in order to construct
  exact solutions. Because the  coupled KdV system is
  obtained by imposing complex variables on the KdV
  equation \eqref{kdv}, we can easily obtain the Darboux transformation for the
  coupled KdV system \eqref{ckdv} from those of the KdV equation.

  It is known that the Lax pair for the  KdV equation \eqref{kdv}
  reads
 \begin{eqnarray}\label{laxx}
  -\Phi_{xx}+U\Phi=\lambda\Phi,
  \end{eqnarray}
  \begin{eqnarray}\label{laxt}
 \Phi_t=-4\Phi_{xxx}+6U\Phi_x+3U_x\Phi.
  \end{eqnarray}
 Substituting $\Phi=\phi_1-i\phi_2$ into Eqs. \eqref{laxx} and \eqref{laxt},
  the Lax pair for Eq. \eqref{ckdv} is as follows:
 \begin{eqnarray}
 &&  \phi_{1xx} = v\phi_2+u\phi_1-\lambda\phi_1,\label{phi1x}\\
 &&  \phi_{2xx}= u\phi_2-v\phi_1-\lambda\phi_2,\label{phi2x}
   \end{eqnarray}
and
   \begin{eqnarray}
 && \phi_{1t}=
  -4\phi_{1xxx}+6v\phi_{2x}+6u\phi_{1x}+3v_x\phi_2+3u_x\phi_1,\label{phi1t}
 \\
 && \phi_{2t}=-4\phi_{2xxx}+6u\phi_{2x}-6v\phi_{1x}+3u_x\phi_2-3v_x\phi_1,\label{phi2t}
  \end{eqnarray}
 then it is easy to prove that the compatibility conditions of four equations
 \eqref{phi1x}--\eqref{phi2t} are exactly the equation system \eqref{ckdv}.

 With the help of the Darboux transformation for the KdV equation \eqref{kdv}, we
 can easily obtain the first step Darboux transformation for the
 coupled KdV system
 \begin{equation}\label{du1}
 u[1]=u-{\ln(f^2+g^2)}_{xx},
 \end{equation}
 \begin{equation}\label{dv1}
  v[1]=v-2{\arctan\left(\frac gf \right)}_{xx},
 \end{equation}
 and the new wave functions are transformed to
 \begin{equation}\label{dphi1}
 \tilde\phi_1=\phi_{1x}-\frac{\ln(f^2+g^2)_x} 2 \phi_1-\arctan\left(\frac
 fg\right)_x\phi_2,
 \end{equation}
\begin{equation}\label{dphi2}
 \tilde\phi_2=\phi_{2x}-\frac{\ln(f^2+g^2)_x} 2 \phi_2+\arctan \left(\frac fg \right)_x\phi_1,
 \end{equation}
 where $\{f, g\}$ is a wave function solution of the Lax pair
 \eqref{phi1x}--\eqref{phi2t} with $\lambda=\lambda_0,\ f=\phi_1,\ g=\phi_2$.
In order to construct the second step Darboux transformation, we
need two wave function seed solutions $\{\phi_{11}, \phi_{12}\}$
and $\{\phi_{21}, \phi_{22}\}$ with two different spectral
parameters $\lambda_1$ and $\lambda_2$ respectively, then the
second step Darboux transformation is
\begin{equation}\label{du2}
u[2]=u-\left[\ln(F^2+G^2)\right]_{xx},
\end{equation}
\begin{equation}\label{dv2}
v[2]=v-2\left[\arctan\left(\frac G F\right)\right]_{xx},
\end{equation}
where two functions $F$ and $G$ are given by
\begin{equation}\label{f}
F=W(\phi_{11},\phi_{21})-W(\phi_{12},\phi_{22}),
\end{equation}
and
\begin{equation}\label{g}
G=-W(\phi_{11},\phi_{22})-W(\phi_{12},\phi_{21}),
\end{equation}
respectively with $W(a,b)=ab_x-ba_x$ is the usual Wronskian
determinant.

 Similarly, the third and fourth step Darboux
transformations are written down as follows simply,
\begin{equation}\label{du3}
u[3]=u-[\ln(H^2+L^2)]_{xx},
\end{equation}
\begin{equation}\label{dv3}
v[3]=v-2\left[\arctan\left(\frac LH \right)\right]_{xx},
\end{equation}
where
\begin{equation*}
H=W(\phi_{11},\phi_{21},\phi_{31})-W(\phi_{11},\phi_{22},\phi_{32})-W(\phi_{12},\phi_{22},
\phi_{31})-W(\phi_{12},\phi_{21},\phi_{32}),
\end{equation*}
and
\begin{equation*}
L=W(\phi_{12},\phi_{22},\phi_{32})-W(\phi_{12},\phi_{21},\phi_{31})-W(\phi_{11},\phi_{22},
\phi_{31})-W(\phi_{11},\phi_{21},\phi_{32})
\end{equation*}
with $\{\phi_{i1},\ \phi_{i2}\} (i=1,2,3)$ are solutions of the
Lax pairs \eqref{phi1x}--\eqref{phi2t} with three different
parameters $\lambda_1, \lambda_2$ and $\lambda_3$. The fourth step
Darboux transformation is more complicated because more wave
functions are included in. The final results reads
\begin{equation}\label{du4}
u[4]=u-[\ln(P^2+Q^2)]_{xx},
\end{equation}
\begin{equation}\label{dv4}
v[4]=v-2\left[\arctan\left(\frac QP \right)\right]_{xx},
\end{equation}
where $P \equiv P(x,t)$ and $Q \equiv Q(x,t)$ are
\begin{eqnarray*}
P(x,t)&=&W(\phi_{11},\phi_{21},\phi_{31},\phi_{41})+W(\phi_{12},\phi_{22},
\phi_{32},\phi_{42})-W(\phi_{12},\phi_{21},\phi_{32},\phi_{41})\nonumber\\
&&-W(\phi_{11},\phi_{21},\phi_{32},\phi_{42})-W(\phi_{12},\phi_{22},\phi_{31},
\phi_{41})-W(\phi_{12},\phi_{21},\phi_{31},\phi_{42})
\nonumber\\
&&-W(\phi_{11},\phi_{22},\phi_{31},\phi_{42})-W(\phi_{11},\phi_{22},\phi_{32},\phi_{41}),
\end{eqnarray*}
and
\begin{eqnarray*}
Q(x,t)&=&W(\phi_{12},\phi_{22},\phi_{32},\phi_{41})+W(\phi_{12},\phi_{21},\phi_{32},
\phi_{42})+W(\phi_{11},\phi_{22},\phi_{32},\phi_{42})\nonumber\\
&&+W(\phi_{12},\phi_{22},\phi_{31},\phi_{42})-W(\phi_{11},\phi_{21},\phi_{32},\phi_{41})
-W(\phi_{11},\phi_{21},\phi_{31},\phi_{42})\nonumber\\
&&-W(\phi_{12},\phi_{21},\phi_{31},\phi_{41})-W(\phi_{11},\phi_{22},\phi_{31},\phi_{41}),
\end{eqnarray*}
respectively, while $\{\phi_{i1}, \phi_{i2}\},\ (i=1,2,3, 4)$ are
four wave function vector of the Lax pair
\eqref{phi1x}--\eqref{phi2t} corresponding to four parameters
$\lambda_{i}, (i=1,2,3,4).$

In general, the $N$-step Darboux transformation for the coupled
KdV system \eqref{ckdv} is given by
\begin{equation}\label{duN}
u[N]=u-[\ln(W_r^2+W_I^2)]_{xx},
\end{equation}
\begin{equation}\label{dvN}
v[N]=v-2\left[\arctan\left(\frac {W_r}{W_I} \right)\right]_{xx}
\end{equation}
with $W_r$ and $W_I$ being the real and imaginary part of the
Wronskian of the $N$ complex wave functions
$$\Phi_i\equiv \phi_{i1}+I\phi_{i2},\ i=1,\ 2,\ ...,\ N,$$
i.e.,
\begin{equation}\label{WN}
W=W(\Phi_1,\ \Phi_2,\ ...,\ \Phi_N)\equiv W_r+I W_I.
\end{equation}

\section{Positon, Negaton and complexiton solutions}

 In this section, we systematically study the positon, negaton and complexiton solutions for
  the coupled KdV system by means of the Darboux transformations given in
  the last section.

  \subsection{Positon solution, $\lambda>0$}

 Based on the first step Darboux transformation, analytical positon solutions
   for the coupled KdV system can be constructed directly.

  Taking the seed solution as $\{u=0, v=0 \}$,
  then solving the Lax pair \eqref{phi1x}-\eqref{phi2t}
  directly, one can find
\begin{eqnarray}\label{phi1}
f&=&\phi_1 = C_2\sin(\sqrt{\lambda} x+4\sqrt{\lambda^3}
t)+C_1\cos(\sqrt{\lambda} x+4\sqrt{\lambda^3} t)\nonumber\\
&=& C_1\cos(\sqrt{\lambda} x+4\sqrt{\lambda^3} t+\delta_1)\equiv
C_1\cos \xi_1,
\end{eqnarray}
\begin{eqnarray}\label{phi2}
g&=&\phi_2 = C_4\sin(\sqrt{\lambda} x+4\sqrt{\lambda^3}
t)+C_3\cos(\sqrt{\lambda} x+4\sqrt{\lambda^3}
t)\nonumber\\
&=& C_3\cos(\sqrt{\lambda} x+4\sqrt{\lambda^3} t+\delta_2)\equiv
C_3\cos \xi_2.
\end{eqnarray}
Substituting the result equations \eqref{phi1} and \eqref{phi2}
into the first step Darboux transformation yields the single
general positon solution for the coupled KdV system \eqref{ckdv}:

\begin{eqnarray}\label{Du11}
u&=&\frac{2\lambda\left[(C_1^2-C_3^2)(C_1^2\cos^2\xi_1-C_3^2\cos^2\xi_2)
    +4C_1^2C_3^2\cos(\delta_1-\delta_2)\cos\xi_1
    \cos\xi_2\right]}{(C_1^2\cos^2\xi_1+C_3^2\cos^2\xi_2)^2},\\
v&=&\frac{4C_1C_3\lambda\left[(C_1^2\cos^2\xi_1-C_3^2\cos^2\xi_2)\cos(\delta_1-\delta_2)
    -(C_1^2-C_3^2)\cos\xi_1
    \cos\xi_2\right]}{(C_1^2\cos^2\xi_1+C_3^2\cos^2\xi_2)^2}.\label{Dv11}
\end{eqnarray}

It is easy to find that the positon solution
\eqref{Du11}--\eqref{Dv11} is always nonsingular for $v\neq 0$.
Actually, from \eqref{Du11}--\eqref{Dv11} we know that the positon
is singular only for
$$\delta_2=\delta_1+n\pi,\ n=0,\ \pm1,\ \pm2,\ ...,$$
while this constant condition leads to $v=0$. This result
coincides with the fact that the positon of the real KdV is
singular.

Fig. 1 shows the nonsingular positon structure
\eqref{Du11}--\eqref{Dv11} with the constant parameter selections
\begin{eqnarray}
 C_1=2,\ C_3=4,\ \delta_1=0,\ \delta_2=1,\ \lambda=4\label{pc1}
\end{eqnarray}
at time $t=0$.

\input epsf
     \begin{figure}
     \epsfxsize=7cm\epsfysize=5cm\epsfbox{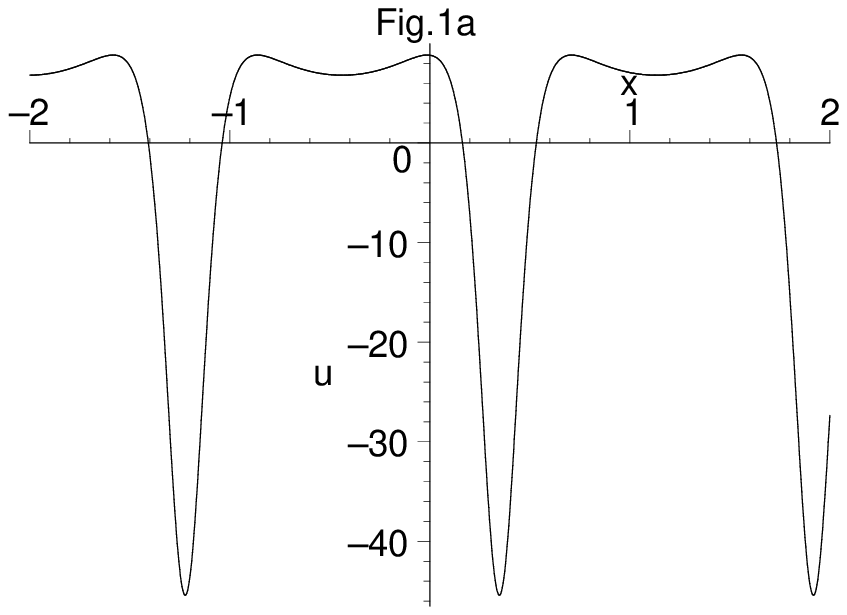}
     \epsfxsize=7cm\epsfysize=5cm\epsfbox{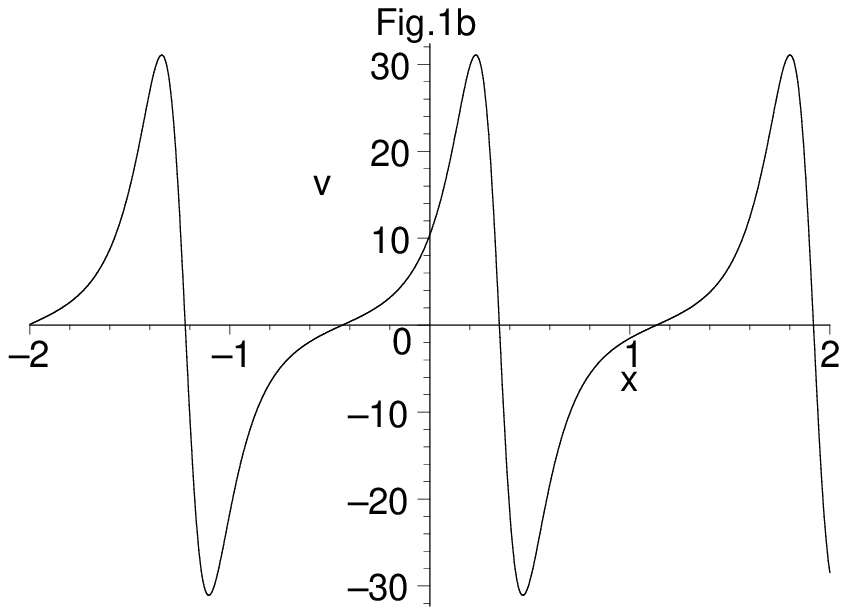}
     \caption{Structur of the Positon solution (a) for the field $u$ expressed by \eqref{Du11} and  (b)
     for the quantity $v$ given by \eqref{Dv11} with the parameter selections \eqref{pc1} at time $t=0$.}
     \end{figure}

From Fig. 1, and the expressions \eqref{Du11} and \eqref{Dv11}, we
know that a nonsingular positon is just a special nonsingular
periodic wave solution.

\subsection{Negaton solution ($\lambda=-k^2<0$)}

In order to obtain the negaton solutions for coupled KdV system
\eqref{ckdv}, we only need to substitute $\lambda=-k^2$ into Eqs.
\eqref{phi1} and \eqref{phi2}, then the wave functions are
\begin{eqnarray}
f&=&C_2 e^{kx-4k^3t}+C_1 e^{-kx+4k^3t}\nonumber\\
&=&c_1\sinh (kx-4k^3t+\Delta_1)\equiv c_1\sinh
\eta_1,\label{sphi1}\\
g&=&C_4 e^{k x-4 k^3 t}+C_3 e^{-k x+4 k^3
t}\nonumber\\
&=&c_2\sinh (kx-4k^3t+\Delta_2)\equiv c_2\sinh \eta_2
\label{sphi2}.
\end{eqnarray}
Substituting the above wave functions into the expressions of the
first step Darboux transformation, we have the single negaton
solution for the coupled KdV system \eqref{ckdv}

\begin{eqnarray}\label{sDu11}
u&=&\frac{2k^2\left[(c_1^2-c_2^2)(c_1^2\sinh^2\eta_1-c_2^2\sinh^2\eta_2)
    +4c_1^2c_2^2\cosh(\Delta_1-\Delta_2)\sinh\eta_1
    \sinh\eta_2\right]}{(c_1^2\sinh^2\eta_1+c_2^2\sinh^2\eta_2)^2},\\
v&=&\frac{2c_1c_2k^2\left[(c_1^2\sinh
(2\eta_1)+c_2^2\sinh(2\eta_2)\right]\sinh(\Delta_1-\Delta_2)
    }{(c_1^2\sinh^2\eta_1+c_2^2\sinh^2\eta_2)^2}.\label{sDv11}
\end{eqnarray}

Fig. 2 shows a special negaton structure under the parameter
selections
\begin{eqnarray}\label{pc2}
c_1=1,\ c_2=5,\ \Delta_1=0,\ \Delta_2=-2,\ k=2
\end{eqnarray}
at time $t=0$.

From the negaton expression \eqref{sDu11}--\eqref{sDv11}, one can
see that the negaton solution of the model is nonsingular except
that $\Delta_2=\Delta_1+n\pi\sqrt{-1}\ (n=0,\ \pm 1,\ \pm2,\ ...)\
$ which is corresponding to the real KdV case, $v=0$.

\input epsf
     \begin{figure}
     \epsfxsize=7cm\epsfysize=5cm\epsfbox{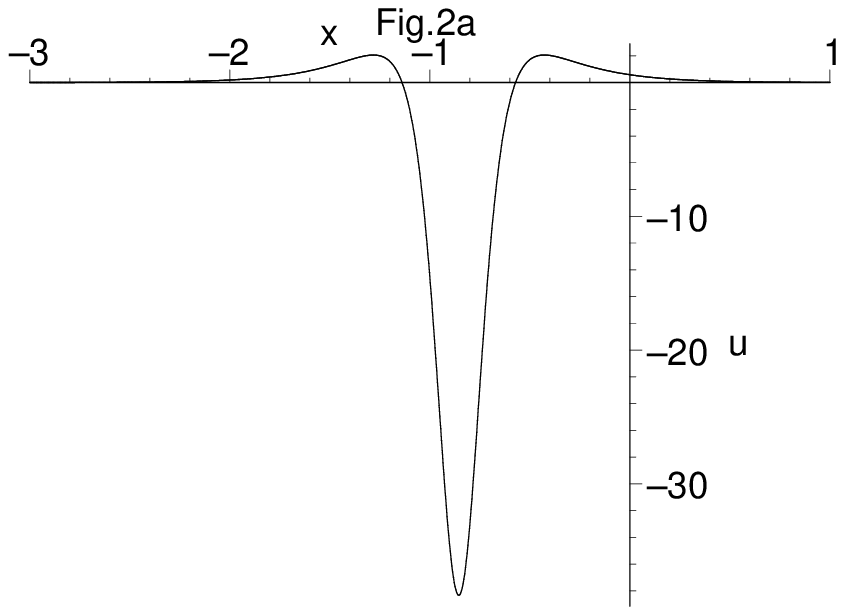}
     \epsfxsize=7cm\epsfysize=5cm\epsfbox{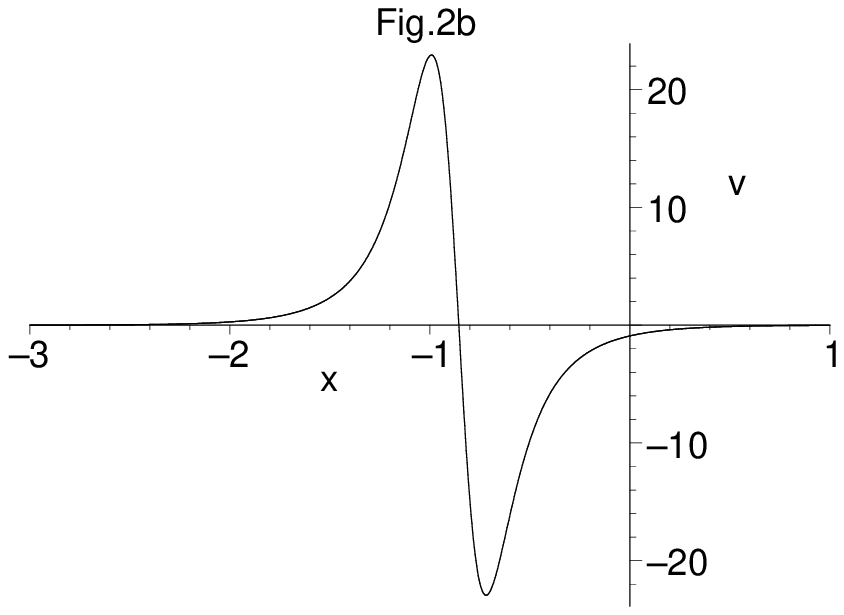}
     \caption{A typical negaton structure of the coupled KdV system \eqref{ckdv} (a) for $u$ expressed by \eqref{sDu11}
     and (b) for $v$ given by \eqref{sDv11} under the parameter selections \eqref{pc2} at time $t=0$.}
     \end{figure}

\subsection{Negaton interaction solutions}

With the N-step Darboux transformations, one may obtain much
richer solution structures including the interactions among
different types of localized excitations. A type of soliton
solution including two negatons can be constructed easily from the
two-step Darboux transformation.
 Let two wave function vectors, $\{ \phi_{11}, \phi_{12}\}$ and $\{\phi_{21}, \phi_{22}
 \}$, are solutions of the Lax pair \eqref{phi1x}--\eqref{phi2t} with the seed $\{u=0,\ v=0\}$ and
 two spectral parameters
 $\{\lambda_1=-k_1^2,
 \lambda_2=-k_2^2\}$ respectively, then we have the two negaton
 solution expressed by \eqref{du2}--\eqref{g} with
\begin{eqnarray}\label{phi11}
&& \phi_{11}=c_1\cosh(k_1x-4k_1^3t+\delta_1),\\
&&\label{phi12}
\phi_{12}=c_2\cosh(k_1x-4k_1^3t+\delta_2),\\
&& \label{phi21}
 \phi_{21}=c_3\cosh(k_2x-4k_2^3t+\delta_3),\\
 &&\label{phi22}
 \phi_{22}=c_4\cosh(k_2x-4k_2^3t+\delta_4).
 \end{eqnarray}

Fig. 3 and Fig.4 display the special two-negaton interaction
procedure for the fields $u$ and $v$ described by
\eqref{du2}--\eqref{g} with zero seed, wave function selections
\eqref{phi11}--\eqref{phi22} and the parameter selections
\begin{eqnarray}\label{pc3}
\delta_1=0,\ c_1=k_1=\delta_2=1,\ k_2=2,\ c_2=\delta_3=3,\
\delta_4=4,\ c_3=c_4=5
 \end{eqnarray}
at times $t=-0.25,\ 0.05,\ 0.1,$ and $0.35$ respectively.
\input epsf
     \begin{figure}
     \epsfxsize=7cm\epsfysize=5cm\epsfbox{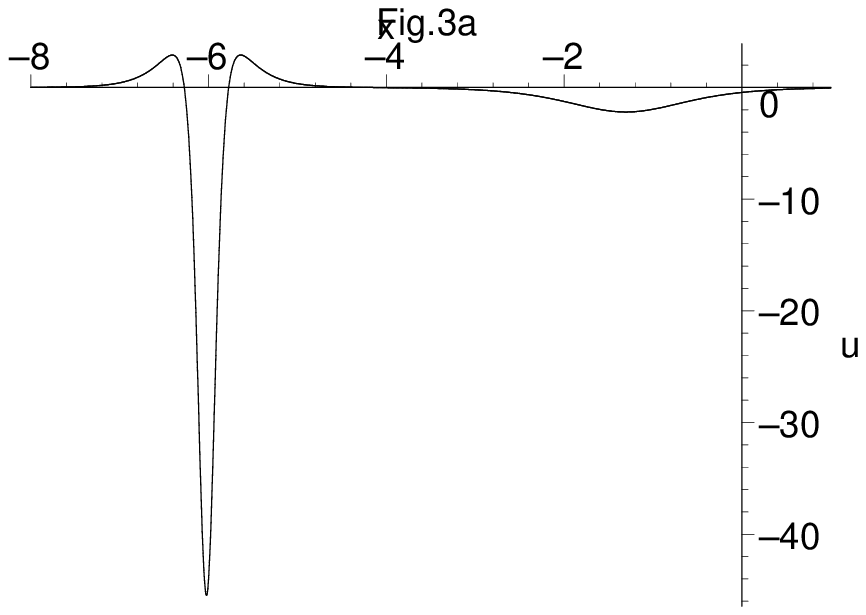}
     \epsfxsize=7cm\epsfysize=5cm\epsfbox{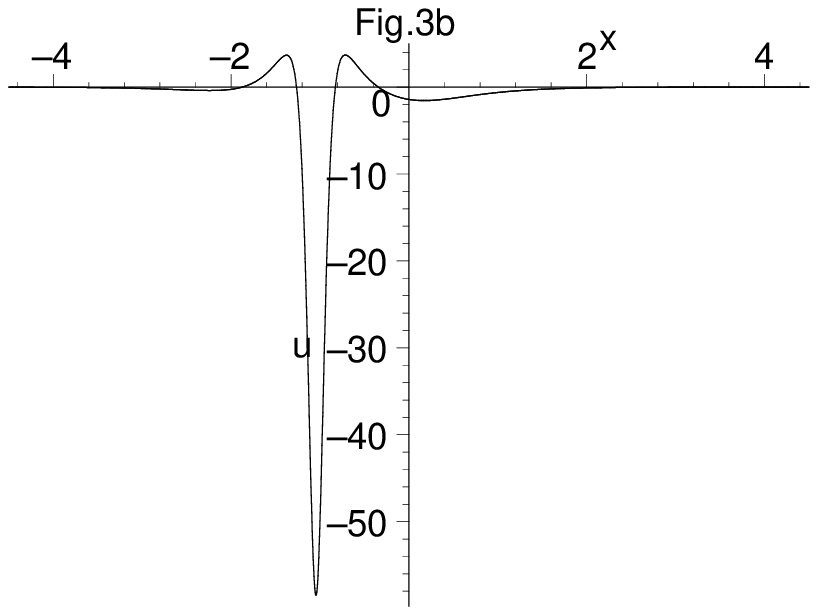}
     \epsfxsize=7cm\epsfysize=5cm\epsfbox{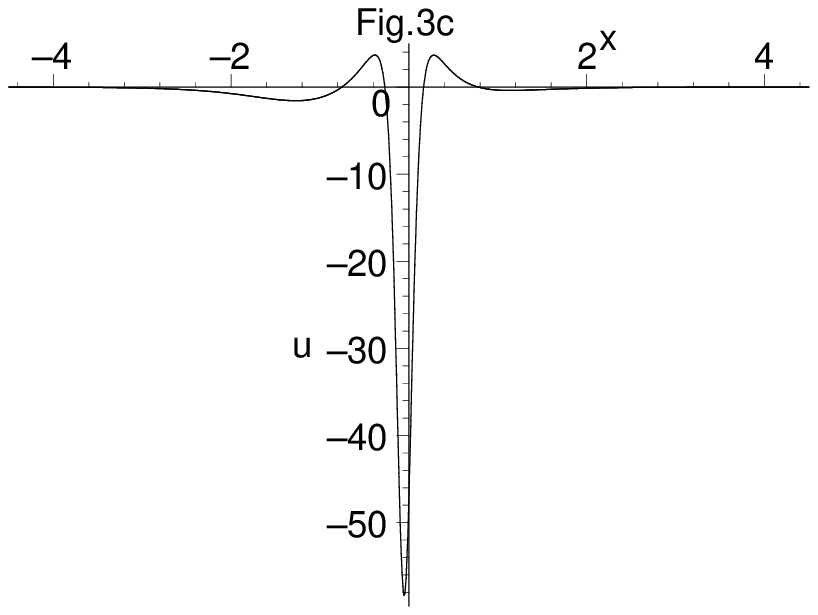}
      \epsfxsize=7cm\epsfysize=5cm\epsfbox{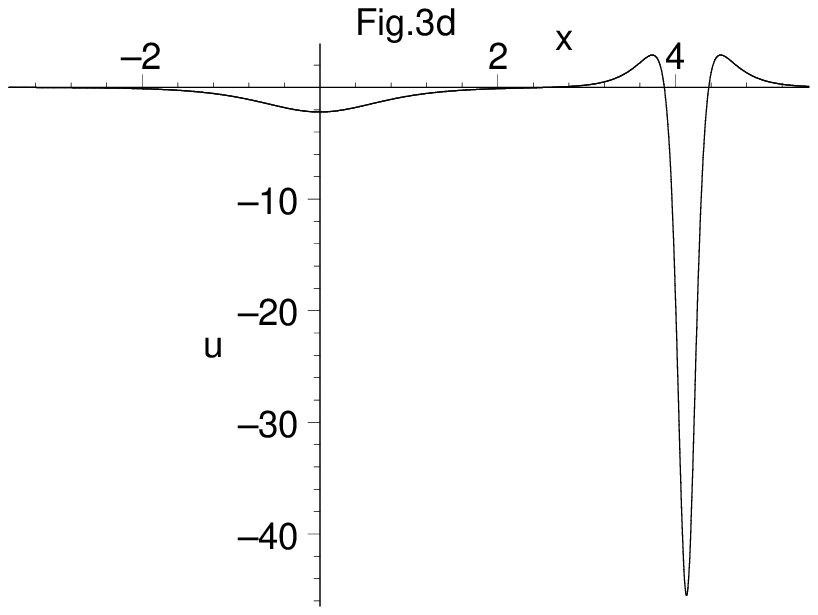}
     \caption{Negaton interaction solution expressed by
     \eqref{du2}--\eqref{g} with zero seed, wave function selections
\eqref{phi11}--\eqref{phi22} and the parameter selections
 \eqref{pc3} for the field $u$ at times (a)$t=-0.25$;
(b) $0.05$; (c) $0.1$ and (d) $0.35$ respectively.}
     \end{figure}

\input epsf
     \begin{figure}
     \epsfxsize=7cm\epsfysize=5cm\epsfbox{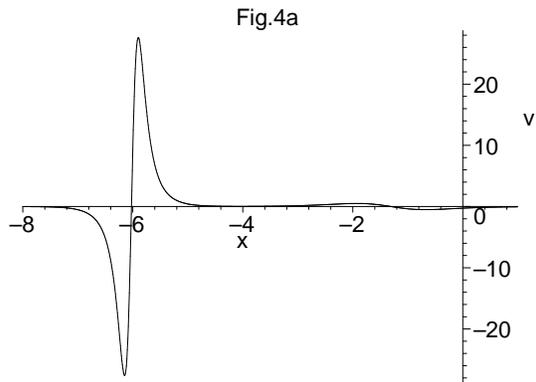}
     \epsfxsize=7cm\epsfysize=5cm\epsfbox{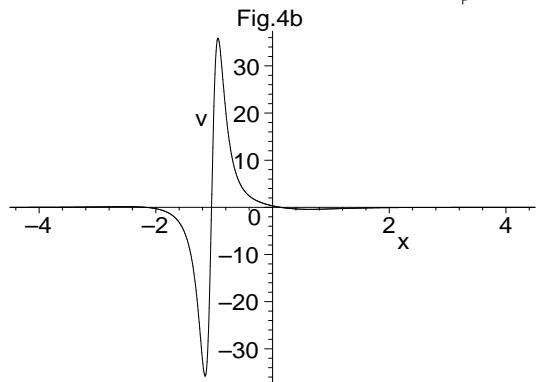}
     \epsfxsize=7cm\epsfysize=5cm\epsfbox{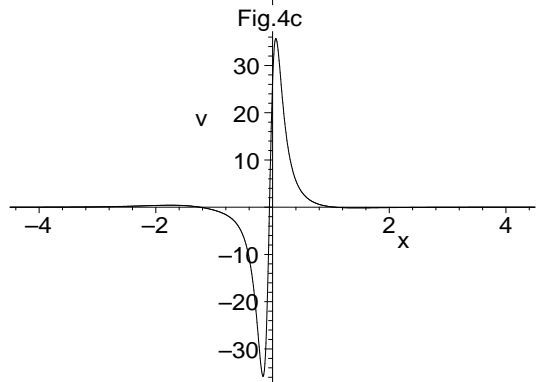}
    \epsfxsize=7cm\epsfysize=5cm\epsfbox{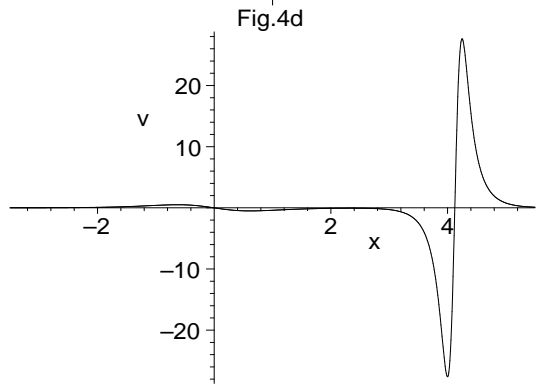}
     \caption{Negaton interaction solution for the quantity $v$ with the same wave function and parameter selections as
     Fig. 3}
     \end{figure}

\subsection{ Positon-negaton interaction solution}

In the same way, by selecting the spectral parameters
appropriately in the N-step Darboux transformations one can obtain
the interaction solutions among positons and negatons. For
instance, based on the second step Darboux transformation by
choosing two different spectral parameters with opposite sign,
$\lambda_1=\kappa^2,\ \lambda_2=-k^2$, then we can obtain the
positon-negaton interaction solutions for the coupled KdV system
\eqref{ckdv}.

To show a concrete example, we further fix the spectral parameters
as $\lambda_1=1$, $\lambda_2=-4$ and select the corresponding
spectral functions in the forms
\begin{eqnarray}\label{pphi11}
\phi_{11}=4\sin(x+4 t)+3\cos(x+4t),
\end{eqnarray}
\begin{eqnarray}\label{pphi12}
\phi_{12}=2\sin(x+4 t)-6\cos(x+4t).
\end{eqnarray}
\begin{eqnarray}\label{pphi21}
\phi_{21}=4 e^{-2x+32t}+ e^{2x-32t},
\end{eqnarray}
\begin{eqnarray}\label{pphi22}
\phi_{22}=4 e^{-2x+32t}+0.01 e^{2x-32t}.
\end{eqnarray}
Substituting \eqref{pphi11}-\eqref{pphi22} and \eqref{f}-\eqref{g}
into \eqref{du2}-\eqref{dv2} with the initial seed solution
$\{u=0, v=0\}$, a special positon-negaton solution follows
immediately. Fig. 6 and Fig. 7 show the interaction procedure
expressed by \eqref{du2}-\eqref{g}, \eqref{pphi11}-\eqref{pphi22}
and the initial solution $\{u=0, v=0\}$ for the fields $u$ and $v$
respectively.

\input epsf
     \begin{figure}
     \epsfxsize=7cm\epsfysize=5cm\epsfbox{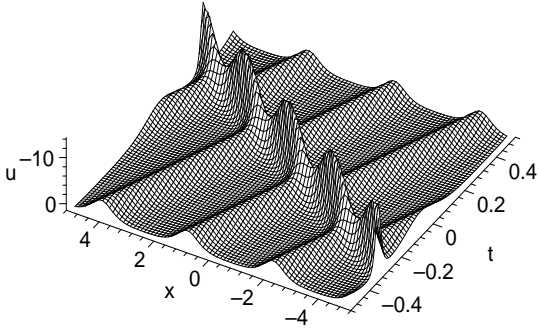}
         \caption{Positon-negaton interaction solution expressed by \eqref{du2}
  and \eqref{f}-\eqref{g}, \eqref{pphi11}-\eqref{pphi22} for the coupled KdV system \eqref{ckdv}.}
     \end{figure}

\input epsf
     \begin{figure}
     \epsfxsize=7cm\epsfysize=5cm\epsfbox{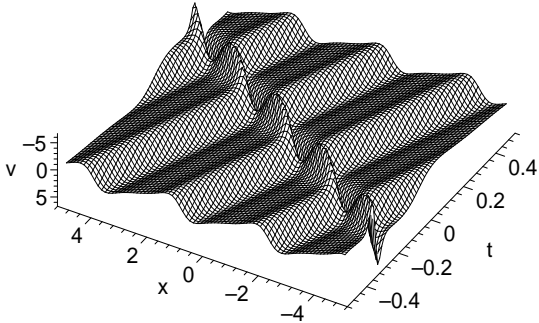}
     \caption{Positon-negaton solution expressed by  \eqref{dv2}-\eqref{g}, \eqref{pphi11}-\eqref{pphi22}
   for the coupled KdV system \eqref{ckdv}.}
     \end{figure}
\subsection{Analytical complexiton solution}
For many integrable system, complexiton solutions have been found
\cite{Ma}. But it should be pointed out that the known complexiton
solutions have singularities. It is lucky that the analytical
complexiton solutions are found in this Letter for the first time.
Let two spectral parameters $\lambda$ and $\lambda'$ be conjugated
complex number, that is $\lambda=\lambda_1+I\lambda_2$ and
$\lambda'=\lambda_1-I\lambda_2$, then we arrive at two coupled
wave solutions with the initial solution $\{u=0, v=0\}$,
\begin{eqnarray}\label{cphi11}
\phi_{11}=IC_1
e^{(4I\lambda_1^3-12\lambda_1^2\lambda_2-12I\lambda_1\lambda_2^2+4\lambda_2^3)t+(\lambda_1
I-\lambda_2)x}+IC_2
e^{(12\lambda_1^2\lambda_2-4I\lambda_1^3+12I\lambda_1\lambda_2^2-4\lambda_2^3)t+(\lambda_2-I\lambda_1)x},
\end{eqnarray}
\begin{eqnarray}\label{cphi12}
\phi_{12}=IC_3
e^{(4I\lambda_1^3-12\lambda_1^2\lambda_2-12I\lambda_1\lambda_2^2+4\lambda_2^3)t+(\lambda_1
I-\lambda_2)x}+IC_4
e^{(12\lambda_1^2\lambda_2-4I\lambda_1^3+12I\lambda_1\lambda_2^2-4\lambda_2^3)t+(\lambda_2-I\lambda_1)x},
\end{eqnarray}
\begin{eqnarray}\label{cphi21}
\phi_{21}=c_1 e^{(4
I\lambda_1^3+12\lambda_1^2\lambda_2-12I\lambda_1\lambda_2^2-4\lambda_2^3)t+(\lambda_1I+\lambda_2)x}+c_2
e^{(-I\lambda_1-\lambda_2)x+(12
I\lambda_1\lambda_2^2+4\lambda_2^3-4I\lambda_1^3-12\lambda_1^2\lambda_2)t},
\end{eqnarray}
\begin{eqnarray}\label{cphi22}
\phi_{22}=c_3 e^{(4
I\lambda_1^3+12\lambda_1^2\lambda_2-12I\lambda_1\lambda_2^2-4\lambda_2^3)t+(\lambda_1I+\lambda_2)x}+c_4
e^{(-I\lambda_1-\lambda_2)x+(12
I\lambda_1\lambda_2^2+4\lambda_2^3-4I\lambda_1^3-12\lambda_1^2\lambda_2)t}.
\end{eqnarray}
Substituting \eqref{cphi11}-\eqref{cphi22} into
\eqref{f}-\eqref{g}, we can obtain
\begin{eqnarray}\label{rf}
F&=&4\lambda_2(c_1C_1-c_3C_3)\sin[2\lambda_1(4t\lambda_1^2+x-12t\lambda_2^2)]\nonumber\\
&&+4\lambda_1(c_4C_3-c_2C_1)\cosh[2\lambda_2(x+12t\lambda_1^2-4t\lambda_2^2)],
\end{eqnarray}
\begin{eqnarray}\label{rg}
G&=&-4\lambda_2(c_3C_1+c_1C_3)\sin[2\lambda_1(4t\lambda_1^2+x-12t\lambda_2^2)]\nonumber\\
&&+4\lambda_1(c_2C_3+c_4C_1)\cosh[2\lambda_2(x+12t\lambda_1^2-4t\lambda_2^2)],
\end{eqnarray}
where $\{c_1, c_2, c_3, c_4, C_1, C_2, C_3, C_4, \lambda_1,
\lambda_2 \}$ are all constants. To show a detailed structure of
the complexiton, we fix the constant parameters further.
 The substitution of \eqref{rf} and \eqref{rg} into
\eqref{du2}-\eqref{dv2} with the fixing constants $\{ c_1 = 5, c_2
= 0, c_3 = 0, c_4=5, C_1 = -1, C_2 =2, C_3=2,
C_4=1,\lambda_1=2,\lambda_2=\frac12\}$, yields the following
analytical nonsingular complexiton solution
\begin{eqnarray}\label{cu2}
U=-\left[\ln(F^2+G^2)\right]_{xx}=\frac{A(x,t)}{B(x,t)}
\end{eqnarray}
and
\begin{eqnarray}\label{cv2}
V=-2\left[\arctan\left(\frac
GF\right)\right]_{xx}=\frac{C(x,t)}{B(x,t)},
\end{eqnarray}
where
\begin{eqnarray*}
 A(x,t)&=&-[272\cos(104 t
+ 8 x) + 240 + 240 \cos(104 t + 8 x) \cosh(2 x + 94 t)\nonumber\\
&&+ 272 \cosh(2 x + 94 t)-128 \sin(104 t + 8 x)\sinh(2 x + 94 t)],
\end{eqnarray*}
\begin{eqnarray*}
B(x,t)&=&\frac{835}8-\frac{17}2\cos(104t+8x)+\frac18\cos(208t+16x)+136\cosh(2x+94t)\nonumber\\
&&-8\cos(104t+8x)\cosh(2x+94t)+32\cosh(4x+188t),
\end{eqnarray*}
and
\begin{eqnarray*}
C(x,t)&=&-[{1862}\sin(52t+4x)\cosh(x+47t)+30\cosh(x+47t)\sin(156t+12x)\nonumber\\
&&+240\cos(52t+4x)\sinh(x+47t)+16\sinh(x+47t)\cos(156t+12x)\nonumber\\
&&+480\sin(52t+4x)\cosh(3x+141t)+256\cos(52t+4x)\sinh(3x+141t)].
\end{eqnarray*}

\input epsf
\begin{figure}
     \epsfxsize=7cm\epsfysize=5cm\epsfbox{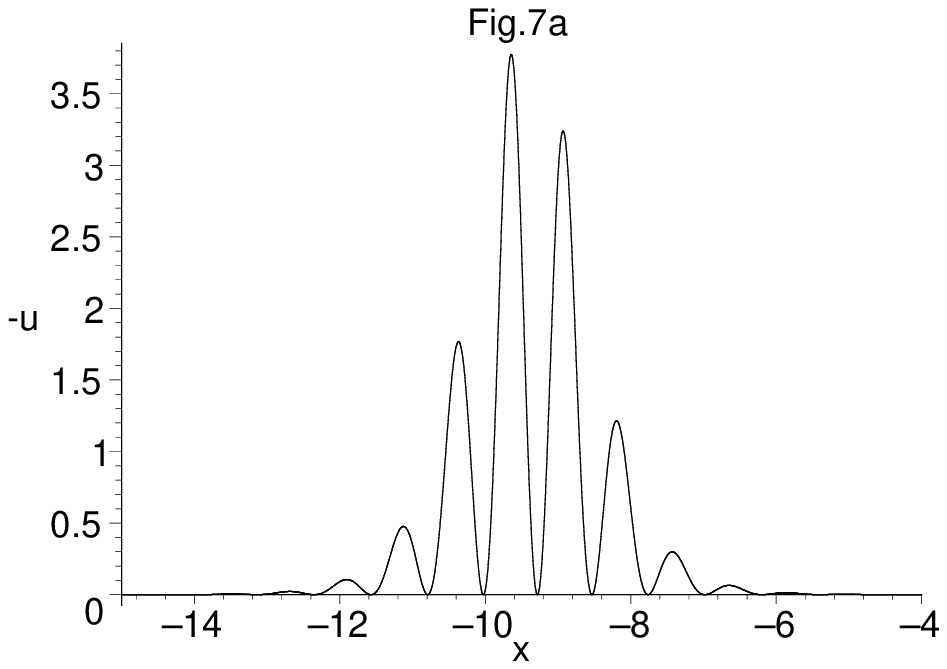}
     \epsfxsize=7cm\epsfysize=5cm\epsfbox{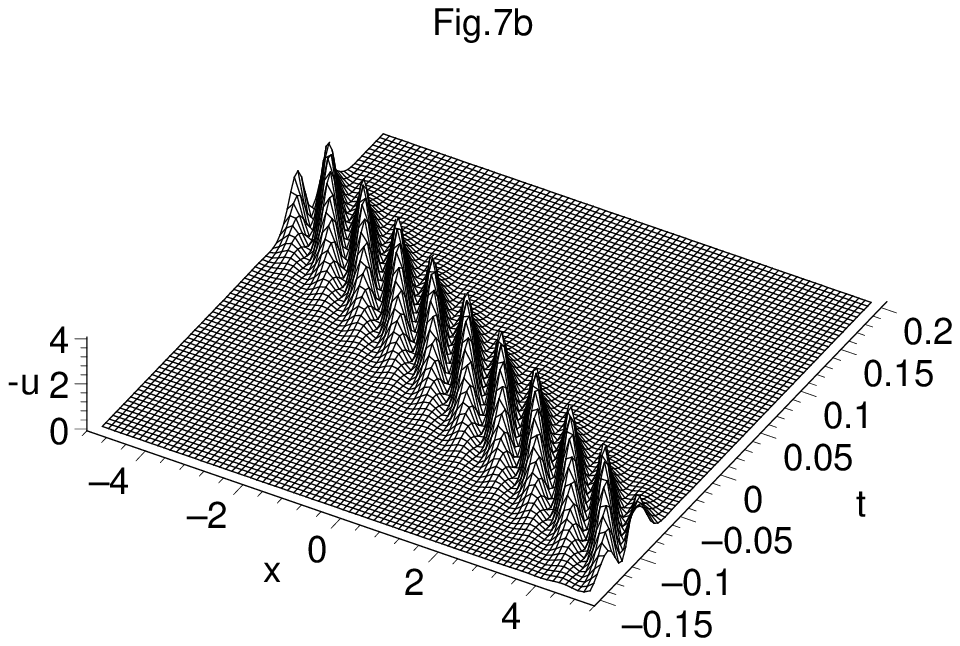}
     \caption{Analytical complexiton solution expressed by
     \eqref{cu2} for the coupled KdV system \eqref{ckdv}.
     (a) The structure plot at the fixed time $t=0.2$; (b) The evolution plot.}
\end{figure}

\input epsf
\begin{figure}
     \epsfxsize=7cm\epsfysize=5cm\epsfbox{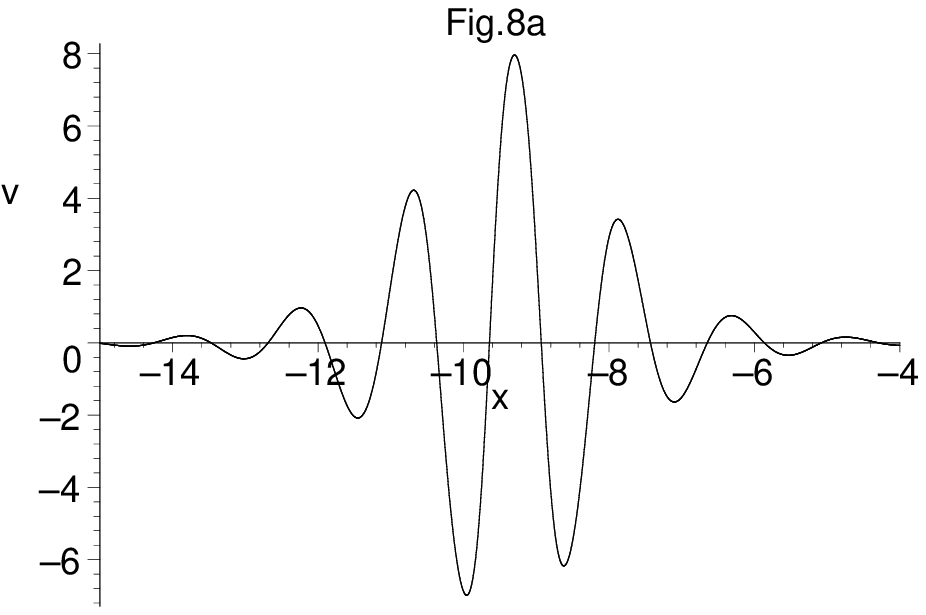}
     \epsfxsize=7cm\epsfysize=5cm\epsfbox{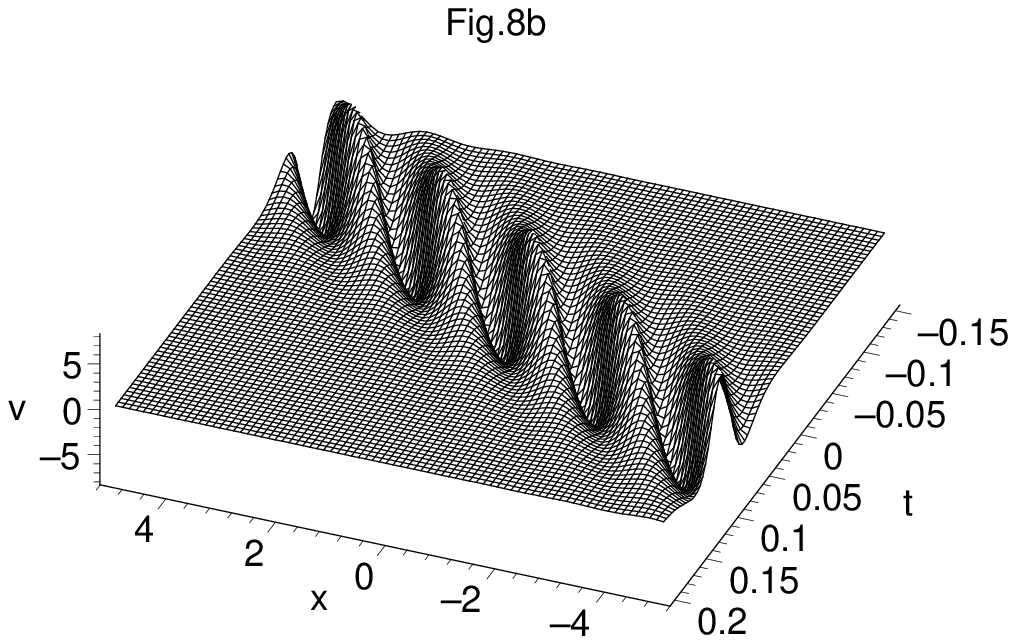}
     \caption{Nonsingular complexiton solution given by
     \eqref{cv2} for the coupled KdV system \eqref{ckdv}.
(a) The structure plot at the fixed time $t=0.2$; (b) The
evolution plot.}
\end{figure}
The complexiton solution displayed above is analytical without any
singularity and it is the first time to find this type of
analytical complexiton solutions for the integrable systems to our
knowledge. Fig. 7 and Fig. 8 show the detailed structures of the
fields $u$ and $v$ expressed by \eqref{cu2} and \eqref{cv2}
respectively.

\section{Summary and discussion }

 In this Letter, exact solutions of a coupled KdV system  which is obtained as a special case
 of the general coupled KdV systems derived from the two
 layer fluid dynamical systems and two-component Bose-Einstein condensates are studied in detail.
The model can also be simply read off from the well known single
KdV equation by assuming its field is complex. Thanks to the model
can be obtained from the complex KdV equation, the Lax pair of the
coupled KdV system follows immediately and so does the Darboux
 transformation.

To study the detailed structures of the localized and periodic
excitations of the model, the first four step Darboux
 transformations are explicitly given in terms of the real Wronskian
 determinant forms while the general N-step Darboux
 transformation is written down with help of the complex Wronskian determinant form.

Starting from the trivial seed solution, the first step of the
Darboux transformation leads to two types of exact solutions, the
negaton and the positon which are related to negative and positive
spectral parameters respectively. Though the single negaton
solution for other integrable models, say, the single real KdV
system, can be both singular and nonsingular, it can be only
analytic for the coupled KdV system \eqref{ckdv}. Similarly, the
positon solution can also be only analytic while for the single
real KdV equation there is no analytic positon.

The solutions obtained from the second step Darboux
 transformation and the trivial zero seed can be two negaton
interaction solution, two positon solution, negaton-positon
solution and a single complexiton dependent on the selections of
the spectral parameters. The interaction among negatons is
completely elastic. The positon-negaton solution can be considered
as a single soliton solution with a periodic wave background. The
single real complexiton solution yields if two spectral parameters
are complex conjugates. The known complexiton solutions for other
integrable systems are singular\cite{Ma}. It is interesting that
the complexiton solution obtained here for the coupled KdV system
is nonsingular! This type of nonsingular complexiton solution may
exist for other types of coupled integrable systems.

Though the coupled KdV system studied here can be read off from
the complex KdV equation, the solution properties are quite
different. For the real single component KdV equation, there is no
nonsingular positons and complexitons. However, for the coupled
KdV system, negatons, positons and complexitons can be
nonsingular.

For the single KdV system, various other types of exact solutions,
such as the rational solutions, cnoidal wave solutions, algebraic
geometry solutions and $\tau$ function solutions have been
obtained by many authors under different approaches. The similar
solutions for the coupled KdV system \eqref{ckdv} can be easily
obtained in some similar ways. However, the detailed properties of
the solutions, especially there analytic behavior, must be quite
different. The more about the model and its possible applications
in fluids, atmospherical dynamics and the Bose-Einstein
condensates are worthy of studying further.

The authors are indebt to thanks Dr. X. Y. Tang and Prof. Y. Chen
for their helpful discussions. The work was supported by the
National Natural Science Foundation of China (No. 90203001 and No.
10475055).

\end{document}